# Enhancing Antimicrobial Molecule Prediction via Dynamic Routing Capsule Networks and Multi-Source Molecular Embeddings

Ruoxi He[1]

**Abstract**

Antibiotics are a vital class of drugs closely associated with the prevention and treatment of bacterial infections. Accurate prediction of molecular antimicrobial activity remains a key challenge in the pursuit of novel antibiotic candidates. However, laboratory-based antimicrobial compounds identification is costly, time-consuming, and prone to rediscovering known antibiotics, highlighting the urgent need for efficient and accurate computational models. Recent advances in machine learning (ML) and deep learning (DL) have significantly enhanced the ability to explore chemical space and identify potential antimicrobial compounds. In this study, we particularly emphasize deep learning models and employ five chemistry language models tailored for chemical data to encode small molecules. Our model incorporates a unique capsule network architecture and introduces innovations in loss function selection and feature processing modules, demonstrating superior performance in predicting inhibitory activities against *Escherichia coli* and *Acinetobacter baumannii*. We conducted a series of ablation studies to elucidate the contributions of network design and input features. Case studies validated the usability and effectiveness of our model.To facilitate accessibility, we developed an intuitive web portal (https://awi.cuhk.edu.cn/~AIDD/CapMolPred/indexpage.php) to disseminate this novel tool. Our results indicate that the proposed approach offers improved predictive accuracy and enhanced interpretability, underscoring the potential of interpretable artificial intelligence methods in accelerating antibiotic discovery and addressing the urgent challenge of antimicrobial resistance.
**Keywords:** antibiotics, antimicrobial activity prediction, bacterial infections, chemistry language models, capsule network, web portal, computational models

## 1 Introduction

Antibiotics are a critical class of drugs primarily used to treat infectious diseases by killing bacteria or inhibiting their growth [1]. Despite the significant success of antibiotics, the increasing prevalence of antibiotic resistance has become a major global health threat, ranking among the top ten health risks according to the World Health Organization (WHO) [2]. In 2019, antibiotic-resistant bacterial infections caused approximately 1.27 million deaths worldwide, with nearly 5 million deaths associated with antibiotic resistance-related diseases. Current trends predict that by 2050, deaths attributable to antibiotic-resistant bacteria will exceed 10 million annually, surpassing cancer mortality rates [3]. Failure to identify risk factors for patients infected with antibiotic-resistant pathogens remains a fundamental cause of inappropriate treatment and subsequent adverse outcomes [4, 5]. The development of novel antibiotics is hindered by multiple factors, including a lack of economic incentives, lengthy and costly research and development cycles, and complex regulatory procedures. Traditionally, antibiotic discovery has relied on screening secondary metabolites from soil microorganisms that inhibit the growth of pathogenic bacteria [6, 7]. This process involves microbial sampling and cultivation, antibacterial activity screening, compound isolation and identification, followed by activity validation and optimization. Although historically successful in the synthesis of important antibiotics such as penicillin and streptomycin, this approach has become less feasible due to its time-consuming nature, high costs, and frequent rediscovery of known compounds [8]. Therefore, early-stage identification of structurally novel candidate molecules is crucial for accelerating development and reducing costs.

The application of machine learning in antibiotic discovery has attracted widespread research interest, enabling rapid exploration of vast chemical and sequence spaces computationally and demonstrating significant advantages in identifying molecules with antibiotic potential[9-14]. Machine learning techniques can accelerate the screening process and substantially reduce the time required for initial preparation [15-18]. With the continuous growth and refinement of biological data, machine learning can efficiently and extensively screen potential antimicrobial molecules. Moreover, employing machine learning to identify antimicrobial compounds structurally distinct from known antibiotics helps avoid redundancy, thereby saving time and resources [19-23].

Recently, Stokes et al.[24] and Liu et al.[25] applied Directed Message Passing Neural Networks (D-MPNN) to predict the antibiotic activity of small molecules and discovered novel compounds with antibacterial activity. Stokes et al. reported the potent antibacterial activity of Halicin, a compound originally developed for treating diabetes [26]. Liu et al. identified Abaucin, an antibacterial compound with narrow-spectrum activity against *A.baumannii*. Mongia et al.[27] employed an Extra Trees/Logistic Regression classifier with $L1$ regularization to link molecular substructures with mechanisms of action in the dataset published by Stokes et al. Lin et al. [28] utilized Graph Convolutional Networks (GCN) [29] combined with attention mechanism modules [30, 31], incorporating functional group features into the input data to achieve enhanced predictive performance. Despite these advances, existing computational methods still exhibit inherent limitations.

Currently, many models utilize RDKit features as molecular descriptors; although computationally efficient, they do not explicitly integrate cross-modal biological context and typically rely on late-stage concatenation to combine features from different modalities. This approach makes it difficult to effectively model synergistic interactions among features. Essentially, such a strategy is static and non-adaptive, assuming all features have equal importance or fixed combination weights in decision-making. Simple concatenated methods cannot dynamically adjust the contribution of different modality features based on the specific context of the input data, potentially overlooking critical nonlinear interactions between modalities, resulting in inadequate information integration or introducing noise. Furthermore, datasets in drug discovery commonly suffer from severe class imbalance. When optimized using unweighted standard cross-entropy loss, prediction performance on minority classes is often poor. Although resampling techniques (oversampling or undersampling) are commonly applied, they may introduce risks of overfitting or information loss. Most existing works rely on graph neural networks such as Chemprop[32] for prediction, which iteratively aggregate neighbor node information to update atom representations. However, final classification depends on global pooling operations, which are constrained by the information loss due to global feature compression.

To address the limitations of existing models, we propose a capsule network (CapsNet) framework[33] incorporating dynamic routing to integrate small-molecule representations derived from multiple large language models (LLMs). As illustrated in Figure 1, the workflow comprises three stages: (1) extracting multi-source molecular embeddings using ChemBERTa-2 [34], Smole-BERT [35], MolFormer [36], MoleBERT [37], and MolCLR [38] ; (2) fusing concatenated features via cross-attention mechanisms; and (3) hierarchically optimizing spatial features through capsule-based dynamic routing. Experimental results demonstrate state-of-the-art performance in antibacterial activity prediction on two real-world datasets.

To summarize, the main contributions are described as follows:

1. Our model demonstrates superior feature extraction capability by integrating multiple chemical-optimized language models (LLMs) pretrained on vast unlabeled compound data, enabling effective capture of complex molecular patterns and chemical nuances. These LLMs generate high-quality representations that accurately reflect underlying bioactive characteristics, allowing meaningful feature extraction from limited labeled datasets for more reliable antimicrobial activity prediction.

2. Our model employs a cross-attention mechanism that dynamically aligns heterogeneous molecular representations through Query-Key-Value computations, preserving topological feature relationships while adaptively focusing on key chemical substructures. This advanced integration approach proves particularly effective for combining diverse LLM-generated molecular embeddings compared to traditional concatenation or pooling methods.

3. Our model innovatively implements capsule networks for molecular representation learning, where vectorized capsules simultaneously encode substructure existence probability (magnitude) and spatial configuration (orientation). The dynamic routing mechanism automatically establishes hierarchical part-whole relationships while maintaining robustness to 3D conformational variations, enabling more precise modeling of molecular interactions.

4. Our model addresses the inherent class imbalance in drug discovery datasets through a novel asymmetric loss function. The loss function implements exponentially amplified gradient updates for active (minority class) samples during backpropagation, with the amplification factor $\gamma$ dynamically adjusted based on real-time batch statistics to prevent overfitting. It also incorporates adaptive class-weight thresholds that automatically recalibrate during training according to the evolving precision-recall characteristics, maintaining a moving equilibrium between false positives and negatives. The loss function's dynamic nature is particularly effective for drug discovery applications where active compounds typically represent only 0.5-5% of screening libraries.

5. Our model features a streamlined end-to-end architecture requiring only molecular SMILES input without pre-computed similarity matrices.

6. Our model was verified using new antibiotics, and the case study further demonstrated the superiority and reliability of our model.

## 2 Related works

This section reviews existing methodologies in machine learning and deep learning in molecular antimicrobial activity, with comprehensive details presented in Table 1.
Stokes et al.[24] employed a directed message passing neural network (Chemprop) to train a D-MPNN model using 2,335 compounds annotated with growth inhibition data. The model predicted molecular properties from graph structures of the *E.coli* dataset generated by RDKit. By applying transfer learning to broad-spectrum prediction tasks, they screened multiple chemical libraries and successfully identified a novel antibacterial molecule named halicin from the Drug Repurposing Hub, along with eight additional antimicrobial compounds exhibiting significant structural divergence from known antibiotics.

Liu et al.[25] employed an ensemble approach combining directed message passing neural networks (Chemprop) with dual-path input architecture (molecular graphs and precomputed features), which was fine-tuned on an experimental Acinetobacter baumannii dataset and ultimately trained on 7,684 experimentally validated compounds. This innovative methodology led to the discovery of abaucin, a narrow-spectrum antibacterial compound specifically active against A. baumannii, along with the elucidation of its mechanism of action involving disruption of lipoprotein transport through targeting LolE[39].

InterPred[27] represents an interpretable prediction framework that employs two distinct classifiers—ExtraTrees and logistic regression, both incorporating L1 regularization—trained to perform multi-label classification using experimental data from seven microbial species: *Escherichia coli, Staphylococcus aureus, Pseudomonas aeruginosa, Acinetobacter baumannii, Candida albicans, Klebsiella pneumoniae*, and *Cryptococcus neoformans*. The model processes ring-based molecular features generated by RDKit, combined with bacterial morphological characteristics derived from Cell Painting imaging data. Through balanced scoring optimization of binary features, the system predicts compounds' mechanisms of action across five target categories (DNA, RNA, protein, cell wall, and membrane targets), subsequently enabling mechanism-based molecular clustering for enhanced interpretability.

MFAGCN[28] introduces an ensemble learning-based graph network framework that incorporates recursive feature elimination (RFE) for feature optimization. Compared to conventional GCN models, MFAGCN innovates through an attention mechanism that dynamically weights information from different neighboring nodes, enabling more nuanced aggregation of molecular substructure information. The model processes comprehensive molecular representations including RDKit-calculated descriptors, multiple fingerprint types (MACCS, ECFP, PubChem), graph-based features, and functional group characteristics. Pretrained on the extensive PubChem antimicrobial activity dataset comprising 12,000 compounds, the framework was subsequently validated on both *E.coli* and *A.baumannii* experimental datasets.

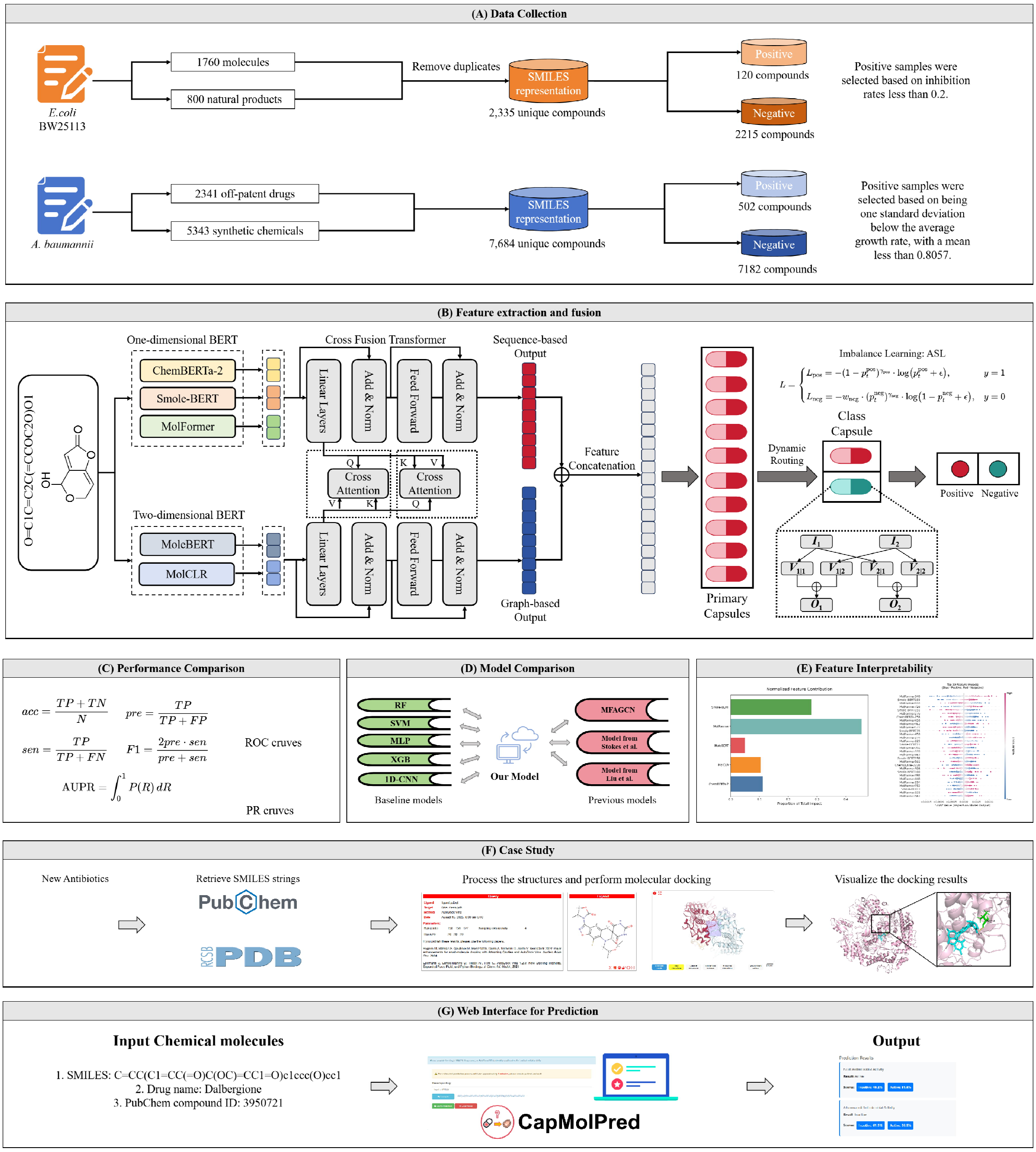

**Figure 1.** This figure illustrates the workflow and application of our research.**(A)** Dataset statistics and binary label generation for *Escherichia coli* and *Acinetobacter baumannii*. **(B)** Workflow illustrating multi-modal feature extraction, subsequent cross attention module, and final prediction utilize a capsule-network predictor with dynamic routing. **(C)** Performance evaluation using standard classification metrics.
**(D)** Comparative analysis against baseline and state-of-the-art models.
**(E)** Feature importance interpretation via SHAP analysis.
**(F)** Case study using newly reported antibiotics compiled from published literature.
**(G)** The model is deployed on a web server, providing a user-friendly interface for predicting antimicrobial activity of molecules.

**Table 1.** The details of existing computational methods.

| Tool | Journal | Dataset | Method Class | Citation | Model Construction | Limitation | Web Interface | Code Available |
|---|---|---|---|---|---|---|---|---|
| Chemprop | Cell (2020) | *E.coli* | Deep Learning (Directed Message Passing Neural Network) | 2454 | A Directed Message Passing Neural Network (D-MPNN) was trained on 2,335 molecules with growth inhibition data, using molecular graph representations as input and applying transfer learning from a broad-spectrum prediction task. | 1. Requires experimental validation for clinical relevance<br>2. Model interpretability challenges for complex molecules | × | https://github.com/chemprop/chemprop |
| Chemprop | *Nature Chemical Biology (2023)* | *A.baumannii* | Deep Learning (Directed Message Passing Neural Network) | 268 | A D-MPNN architecture was fine-tuned for A. baumannii using dual-path inputs—molecular graphs and precomputed features—and trained it on 7,684 experimentally tested compounds, leveraging transfer learning from a broad-spectrum antibiotic model. | 1. Limited training data for narrow-spectrum antibiotics<br>2. Computational screening bias toward synthetic feasibility | × | https://github.com/chemprop/chemprop |
| InterPred | Science Advances (2019) | *E.coli* & *A.baumannii* | Interpretable Machine Learning (Multi-task Classification) | 13 | Extra trees/logistic regression classifier with L1 regularization is trained on bacterial morphological features from Cell Painting data to predict five classes of mechanism of action (DNA, RNA, protein, cell wall, and membrane targets), leveraging transfer learning from antibiotic resistance data. | 1. Limited to Gram-negative bacteria *(E. coli* focus)<br>2. Morphological features require high-content imaging<br>3. Cannot detect novel MoA outside training classes | × | https://gitlab.com/mongolicious/interpretableml-for-mechanism-of-action/ |

| | | | | | | | |
|---|---|---|---|---|---|---|---|
| MFAGCN | Scientific Reports (2024) | Machine Learning (Ensemble Learning ) | 1 | Molecular descriptors (209-D) were calculated using RDKit, followed by feature selection via Recursive Feature Elimination (RFE). An XGBoost classifier was trained on the PubChem antimicrobial activity dataset containing over 12,000 compounds. | 1. Dependency on existing activity labels (potential labeling bias)<br>2. Limited generalizability to novel structural domains | × | https://github.com/MFAGCN/MFAGCN. |
| our model | - | Capsule Network | - | A multi-modal model was constructed by extracting features from molecular representations using both 1D sequence encoders (ChemBERTa-2, SmoleBERT, MolFormer) and 2D graph encoders (MoleBERT, MolCLR). These features were fused through a cross-attention module combined with concatenation, and the final prediction was made using a capsule network architecture with dynamic routing. | - | √ | - |

*The citation counts in this table were retrieved on Jul 15, 2025 from Google Scholar.

## 3 Materials and methods

We employed two distinct datasets for model training and evaluation, with their key characteristics summarized in Table 2. The datasets were specifically curated to ensure a comprehensive assessment of our model's performance across different biochemical contexts.

**Table 2.** Statistics of sample distribution in *E.coli* and *A.baumannii* datasets.

| Dataset | Positive sample | Negative sample | Total |
|---|---|---|---|
| *E.coli* | 120 | 2215 | 2335 |
| *A.baumannii* | 502 | 7182 | 7684 |

### 3.1 Datasets

This study selected *Escherichia coli* and *Acinetobacter baumannii* as the core research strains, mainly based on the following scientific considerations: *E.coli*, as a Gram-negative type strain, provides an ideal model for high-throughput screening of antibacterial drugs with its standardized culture system and clear genetic background. The choice of *A.Baumannii* stems from its increasingly severe clinical threat, especially in patients with weakened immune systems, leading to serious complications and high mortality rates. By conducting parallel research on these two pathogens, which respectively represent the classical model bacteria and the emerging threat of drug resistance, not only can the universality of the antibacterial strategy be verified, but also the urgent clinical needs can be specifically addressed.

The first dataset[24] contains growth inhibition data for *E.coli* BW25113, comprising 1,760 structurally diverse synthetic molecules from the widely available US Food and Drug Administration (FDA)-approved drug library[40], and 800 natural products derived from plants, animals, and microorganisms. After removing duplicates, the final dataset includes 2,335 unique compounds. Each compound is provided with standardized SMILES (Simplified Molecular Input Line Entry System) representations and corresponding growth inhibition rates against BW25113. Using 80% growth inhibition as the activity threshold, the primary screening identified 120 molecules exhibiting significant growth inhibitory activity against *E. coli*.

The second dataset[25] in our study specifically examines growth inhibition profiles of *A.baumannii*, a clinically prevalent Gram-negative pathogen notorious for its escalating role in hospital-acquired infections and rapid development of antibiotic resistance. This dataset aggregates compounds from multiple high-throughput screening sublibraries at the Broad Institute, comprising 2,341 off-patent drug molecules and 5,343 synthetic chemicals (totaling 7,684 compounds), each with standardized SMILES[41] representations and corresponding growth inhibition rates. Using statistical analysis, we established a binary classification criterion based on one standard deviation below the mean growth rate (threshold=0.8057) to determine antimicrobial activity.

To address class imbalance during model training, we employ an asymmetric loss function[42] specifically designed to mitigate positive-negative sample imbalance in multi-label classification through differential modulation factors. The core formulation of this loss function is defined as:

$$\mathscr{L}_{\mathrm{ASL}}=\begin{cases} L_{+}=(1-p)^{\gamma_{+}}\log(p) \\ L_{-}=(p_m)^{\gamma_{-}}\log(1-p_m) \end{cases} \quad (1)$$

The range of p is between 0 and 1, which is the probability offset correction for negative samples, $\gamma_{+}$ and $\gamma_{-}$ are the focusing parameters for positive and negative samples respectively. We adopt differentiated focusing parameter configurations (detailed in Supplementary Table 1) according to dataset-specific class distributions. This function reduces the weight of easily classifiable samples while suppressing simple negative samples and samples with probabilities lower than the threshold *m*. Unlike the traditional focal loss function, when the positive samples are smaller than the negative ones, we do not simply increase or decrease the weights of the samples. This scheme is usually set. By decoupling the decay rates ($\gamma+ \neq \gamma-$) of positive and negative samples, it emphasizes the contribution of positive samples while avoiding the problem of overcompensation for rare positive samples. This asymmetric focusing mechanism ensures that the network can not only extract discriminative features from sparse positive samples but also effectively suppress the interference of noisy negative samples.

### 3.2 **Architectural Overview**

The model architecture follows a three-stage cascade design: (1) SMILES feature extraction: SMILES strings are transformed into dense vector representations using pretrained chemical language models; (2) Feature refinement: A cross-attention mechanism dynamically reweights sequence features to generate spatially-aware optimized features; (3) Activity prediction: A capsule network with dynamic routing captures high-order relationships between molecular structures and antibacterial activity, ultimately producing binary classification results. This hierarchical feature distillation framework enhances the model's capability to identify rare active molecules.

#### 3.2.1 Procedure 1: Extracting features from five different BERT model

While traditional molecular representation methods (e.g., one-hot encoding[rodriguez] and molecular fingerprints) are computationally simple, their non-contextual embedding nature fails to capture continuous semantic patterns in molecular sequences. To overcome this limitation, our study employs five BERT-based pretrained molecular representation models (Figure 1B) for contextual feature extraction and integration from SMILES strings. As linearized representations of chemical molecules, SMILES strings encode comprehensive structural information, including atom types, bond configurations (single/double/aromatic bonds), ring systems, and stereochemical features. The pretrained models enable cross-dimensional representation learning from symbolic sequences to 3D chemical properties through deep semantic understanding.

This study employs two categories of molecular representation models for comparative analysis: For sequence modeling, we utilize three Transformer-based pretrained models (ChemBERTa-2, Smole-BERT, and MolFormer) that effectively capture contextual dependencies and long-range chemical patterns in SMILES strings through their self-attention mechanisms; for graph modeling, we implement MoleBERT and MolCLR which convert RDKit molecular objects into graph Data objects compatible with PyTorch Geometric (with simplified atom/bond features represented as indices), enabling joint encoding of molecular graph topology and sequence information to preserve both structural connectivity and chemical attributes. These two approaches provide complementary technical pathways for molecular representation learning from sequence and graph perspectives.

3.2.2 Procedure 2: Applying a cross-attention module to the concatenated features

The cross-attention mechanism is an advanced variant of attention in deep learning, specifically designed to model interactive relationships between different modalities or sequences. Its core principle involves dynamically computing association weights between source and target modalities to achieve targeted information filtering and fusion. In this study, building upon the selected upstream feature extraction models, we implement a dual-stream attention mechanism that incorporates both intra-modal self-attention (sequence↔sequence, graph↔graph) and cross-modal attention (bidirectional sequence↔graph interaction). The key innovation of cross-attention, compared to conventional self-attention, lies in its heterogeneous Query-Key design - using target modality features as Queries and source modality features as Keys/Values - thereby generating joint representations with cross-modal consistency. The computational process can be formally expressed as:

$$\mathrm{CrossAttention}(Q_{\mathrm{seq}},K_{\mathrm{graph}},V_{\mathrm{graph}})=\mathrm{softmax}\left(\frac{Q_{\mathrm{seq}}K_{\mathrm{graph}}^{\top}}{\sqrt{d_k}}\right)V_{\mathrm{graph}} \quad (2)$$

Among them, Q/K/V come from different modalities respectively and $\sqrt{d_k}$ is the scaling factors, where $d_k$ is the dimension of the Key vector. Its core function is to adjust the magnitude of the attention score to ensure the stability of the gradient and the convergence of the model.

3.2.3 Procedure 3: Predicting molecular antimicrobial activity using capsule network

This study innovatively integrates hybrid feature embeddings into a dual-level capsule network architecture to enhance antibacterial activity prediction performance. The network extracts broad chemical interaction features (e.g., hydrogen bonding, hydrophobic interactions as basic representations) through the primary capsule layer, while refining specific antibacterial-related patterns via the dynamic routing mechanism in the class capsule layer, establishing a complementary hierarchical feature learning system.

The hybrid feature embedding constructs joint representations by fusing SMILES sequence features with molecular graph topological information. The output layer generates probability scores in the 0-1 range, directly quantifying antibacterial activity confidence. This dual-capsule approach significantly improves prediction accuracy. Unlike traditional neural networks where max-pooling layers discard spatial information by retaining only the most active neurons in local receptive fields, the capsule network with dynamic routing successfully overcomes this limitation.

Capsule networks employ vectorized representations, where the vector magnitude encodes entity existence probability, and the directional vector integrates multidimensional feature attributes. This vector-based representation enables richer and more refined encoding of pattern features, highlighting the innovative approach of CapsNets in addressing the shortcomings of previous neural network models.

The structural characteristic of capsule networks lies in their hierarchical arrangement of capsules, which mirrors traditional neural layers yet differs significantly in architectural and functional complexity. During the construction of the primary capsule layer, we implemented a 1D convolutional layer with a kernel size of 10, which partitions the feature maps into a series of 8-dimensional capsules. This layer serves dual purposes: extracting salient features and preparing for subsequent dynamic routing.

Furthermore, this stage incorporates a novel nonlinear activation mechanism termed the "squash" function. Specifically designed to modulate the magnitude of output vectors while preserving their directional information, this function is mathematically formulated as:

$$squash(s)=\frac{||s||^2}{1+||s||^2}\frac{s}{||s||} \quad (3)$$

The length of the original vector $\frac{s}{||s||}$ is scaled to $\frac{||s||^2}{1+||s||^2}$. Thus, by shrinking the output vector length to between zero and one, the model can maintain the probability interpretation.

In binary classification scenarios, the class capsule layer architecture employs two 16-D vector representations, corresponding to positive and negative classification outcomes respectively, both encoding structural information from input sequences of molecules. The predictive vector transformation between capsules i and j involves two computational stages: (1) linear projection of primary capsule outputs through a trainable parameter matrix $W_{ij}$, followed by (2) $S_j$ determination via weighted summation of all computed $\hat{u}_{(j|i)}$., where the weighting coefficients are dynamically optimized during the routing-by-agreement process. This formulation preserves the hierarchical feature abstraction capability while ensuring differentiable parameter learning.

$$\hat{u}_{(j|i)}=W_{ij}u_i \quad (4)$$

$$S_j=\sum_{i=1}^{L} c_{ij}\,\hat{u}_{(j|i)} \quad (5)$$

The coupling coefficients $c_{ij}$ are determined through a correlation analysis between the predicted vectors $\hat{u}_{(j|i)}$ from primary capsules and the actual outputs $v_j$ of class capsules during dynamic routing. These coefficients establish a probabilistic distribution characterizing the connection strengths between capsule layers, where the magnitude reflects the activation contribution from specific target sequences and molecular structures, with L denoting the total number of primary capsules.

The dynamic routing algorithm operates through an iterative refinement process that takes three key parameters as input: the prediction vector $\hat{u}_{(j|i)}$, the predefined number of routing iterations $r$ , and the primary capsule layer $l$ . The procedure initiates by initializing all routing logits $b_{ij}$ to zero for every capsule pair (i,j), where $i \in l$ and $j \in l{+}1$. During each iteration cycle, the algorithm first computes coupling coefficients $c_{ij}$ through softmax normalization of $b_{ij}$, then aggregates weighted predictions to determine the intermediate state $S_j$ for each higher-level capsule $j$. This state is subsequently transformed into the output vector vj via the non-linear squashing function. The routing logits are then updated $b_{ij}$ by adding original $b_{ij}$ and $\hat{u}^T_{(j|i)} v_j$ to enhance agreement between connected capsules. This cyclic process continues for $r$ iterations before returning the final capsule outputs $v_j$ , which progressively improve in their capacity to represent hierarchical relationships through each update.

The scalar product $b_{ij} = \hat{u}^T_{(j|i)} v_j$ serves as the log prior probability between the primary capsule i and the class capsule j. Consequently, the sum of the coupling coefficients from the primary capsule i to the class capsules, $\sum_{k=1}^{N} c_{ik}$, equals 1, where N stands for the number of class capsules. The iteration number $r$ is a hyper-parameter that must be predetermined. During dynamic routing, the class capsule layer generates two output vectors $v_j$ . The elements within $v_j$ encode the features, and their length indicates the probability distribution between the two types, positive or negative. Thus, the final section of the network focuses on computing this length, calculated as follows:

The scalar product $b_{ij} = \hat{u}^T_{(j|i)} v_j$ represents the log prior probability quantifying the agreement between primary capsule i and class capsule j. This formulation ensures the coupling coefficients cij from each primary capsule i maintain unit normalization ($\sum_{k=1}^{N} c_{ik}$, where N denotes the class capsule count), achieved through softmax transformation. The iteration count operates as a critical hyperparameter requiring explicit predefinition. Through dynamic routing, the class capsule layer produces dual output vectors $v^+$ and $v^-$, whose directional components encapsulate feature representations while their magnitudes $\|v_j\|$ specify categorical probabilities (positive vs negative classification). The network's terminal computation consequently focuses on these magnitude determinations, implemented via the squashing function:

$$p_j = \|v_j\|_2 \quad (6)$$

In the architecture of the CapsNet, each capsule $c$ within the final layer is associated with a designated loss function $L_c$ as follows [43]:

$$L_c=T_c\max(0,m^{\pm}//v_c//)^2+\lambda\left(1-T_c\max(0,//v_c//-m^{-})^2\right. \quad (7)$$

where $l$ denotes the classification category and $L_l$ indicates whether a sample belongs to class $l$ (1 if true, 0 otherwise).

## 3.3 Evaluation

In binary classification tasks based on deep learning, model performance is typically evaluated using metrics such as accuracy, specificity, sensitivity, precision, and the F1-score, which are defined as follows:

$$Accuracy=\frac{TP+TN}{P+N} \quad (8)$$

$$Sensitivity=\frac{TP}{TP+FN} \quad (9)$$

$$Specificity=\frac{TN}{TN+FP} \quad (10)$$

$$Precision=\frac{TP}{P} \quad (11)$$

$$F1\ score=\frac{2Sensitivity * Precision}{Sensitivity+Precision} \quad (12)$$

$$MCC=\frac{TP\times TN-FP\times FN}{\sqrt{(TP+FP)(TP+FN)(TN+FP)(TN+FN)}}$$

where TP is true positive, TN is true negative, FP is false positive, FN is false negative, T is positive, and N is negative.

In antibiotic discovery research, datasets commonly exhibit class imbalance, characterized by a substantial predominance of negative samples (non-antibiotic molecules) over positive samples (known antibiotic molecules). This imbalanced distribution renders the area under the ROC curve (AUC-ROC) potentially inadequate for proper model evaluation. Under such circumstances, the area under the precision-recall curve (AUPRC) emerges as a more suitable metric due to its sensitivity to minority class samples. The AUPRC quantifies model performance in identifying positive samples (antibiotic molecules) by calculating the integral area under the precision-recall curve. Our findings demonstrate that higher AUPRC values correlate with superior model performance in detecting molecules with antibacterial activity. Employing AUPRC as the primary evaluation metric enables effective screening of high-potential antibacterial candidates, thereby significantly reducing the workload and costs associated with subsequent experimental validation.

## 4 Results

The dataset was split into training and test sets at an 8:2 ratio. The model was developed using the TensorFlow and Keras frameworks, with the following training protocols. Early stopping was applied to halt training if no improvement in validation accuracy was observed over 50 consecutive epochs, preventing the model from overfitting. The Adam optimizer was employed with a batch size of 64 and an initial learning rate of 0.0001.

### 4.1 Performance of our model compared with previous models

To comprehensively evaluate model performance across both bacterial strains *E.coli* and *A.baumannii*, we conducted a comparative analysis of current machine learning and deep learning approaches relevant to molecular antibiotic activity prediction in our study.

**Model from Stokes et.al. (2020) .**
A directed-message passing neural network (Chemprop) predict molecular properties from RDKit-generated graph structures on the *E.coli* dataset. Transfer learning was applied from broad-spectrum prediction tasks to train a Directed Message Passing Neural Network (D-MPNN) on 2,335 compounds annotated with growth inhibition data.

**Model from Liu et.al. (2023).**
A directed-message passing neural network (Chemprop) utilized an *Acinetobacter baumannii* experimental dataset and fine-tuned with dual-path inputs (molecular graphs and precomputed features). Transfer learning was applied using a pre-trained broad-spectrum antibiotic model, and the final training was conducted on 7,684 experimentally tested compounds.

**InterPred (2022).**
Both ExtraTrees/logistic regression classifier with L1 regularzation is trained to perform multi-label classification on experimental data from *Escherichia coli*, *Staphylococcus aureus*, *Pseudomonas aeruginosa*, *Acinetobacter baumannii*, *Candida albicans*, *Klebsiella pneumoniae*, and *Cryptococcus neoformans*, combined with bacterial morphological features from Cell Painting. Binary features were optimized via balanced scoring to predict five mechanisms of action (DNA, RNA, protein, cell wall, and membrane targets).

**MFAGCN (2024).**
An ensemble learning-based graph network framework was developed with recursive feature elimination (RFE) for feature optimization. Molecular descriptors were calculated using RDKit, and multiple molecular fingerprints (MACCS, ECFP, PubChem), graph representations, and functional group features were integrated. The model was pre-trained on the PubChem antimicrobial activity dataset (containing >12,000 compounds) and subsequently validated on *E.coli* and *A.baumannii* datasets.

Our model exhibits outstanding performance compared to other existing approaches for predicting antimicrobial activity, as comprehensively summarized in Figure 2A. The detailed indicators are presented in Table S1. For all the studies, we only compared the models that were open-source and could be directly accessed at the time of conducting this research. For the studies by Stokes et al. and Liu et al., the original literature only provided the AUCROC indicator. We trained and tested using the dataset we divided according to the method mentioned in their papers, and the results are shown in the table. On the *E.coli* dataset, our model achieves an impressive AUC-ROC of 0.918, surpassing the performance of Stokes et al.'s model and showing noticeable improvement over MFAGCN. Similar superior performance is observed on the *A.baumannii* dataset, where our model attains an AUC-ROC of 0.781, outperforming Liu et al.'s approach. Notably, while MFAGCN also evaluated both datasets using AUPR as the evaluation metric to account for imbalanced class distributions, our model demonstrates consistently better performance across all metrics, with more comprehensive and reliable results. The consistent advantages observed across all key evaluation metrics, including accuracy, F1 score, AUC, and AUPR, collectively demonstrate that our model provides more balanced and comprehensive improvements, as clearly evidenced by the experimental results.

4.2 Effectiveness of CapsNet

The performance evaluation in Figure 2B demonstrates the significant advantages of our proposed CapsNet model over conventional approaches using both *E. coli* and *A. baumannii* datasets. The AUC indicator is presented in the bar chart. Detailed indicators are shown in Table S2. For the *E. coli* dataset, CapsNet achieved improved accuracy from 0.968 to 0.979, a remarkable 59.9% increase in sensitivity from 0.417 to 0.667, and 33.5% enhancement in F1-score from 0.571 to 0.762, while maintaining comparable specificity. More importantly, the substantial improvements in both AUCROC (0.894→0.918) and AUPR (0.635→0.794, +25.0%) metrics clearly demonstrate CapsNet's superior class discrimination capability and better precision-recall balance, consistently indicating its outstanding performance in antimicrobial activity prediction tasks.

The proposed CapsNet model also demonstrated significant advantages in the *A. baumannii* dataset. While the metric distributions were imbalanced, CapsNet maintained accuracy and specificity within 1% of XGBoost while substantially improving sensitivity from 0.190 to 0.280 and F1-score from 0.309 to 0.392, together with higher AUC and AUPR scores. This indicates that in terms of overall performance, the CapsNet model maintains extremely high specificity while having better robustness in imbalanced scenarios compared to 1D-CNN, providing higher confidence in positive predictions. Overall, our model is the optimal combination of consistency indicators.

To further validate the model's effectiveness, we employed t-SNE visualization to analyze the feature extraction capability. As shown in Figures 2E and 2F, positive (yellow dots) and negative (blue dots) samples exhibited clearer separation in the feature space, providing visual evidence that CapsNet can more effectively capture discriminative features in both datasets, thereby enhancing classification performance.

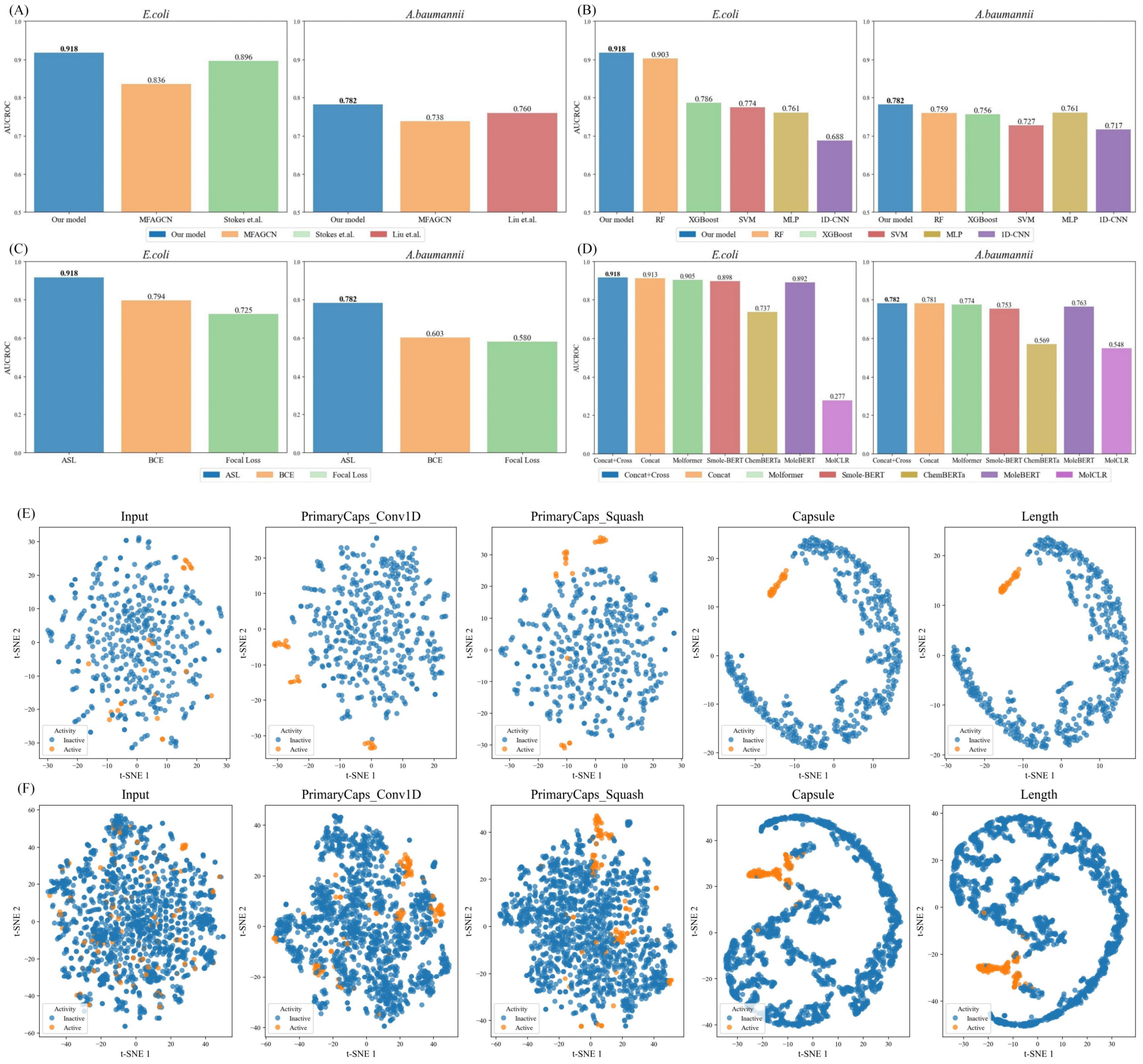

**Figure 2.** Consolidated comparison of model performance and feature-space visualization across *E. coli* and *A. baumannii* datasets. Panels: (A) Comparative binary-classification performance of our proposed model versus existing methods on *E. coli* and *A. baumannii*; (B) Ablation study of different network classifier for both datasets; (C) Comparative performance using different loss functions for both datasets; (D) Performance across alternative feature representations for both datasets; (E-F) t-SNE visualizations of selected CapsNet layers for *E. coli* (E) and *A. baumannii* (F), with points colored by class label.

### 4.3 Effectiveness of Asymmetric Loss Funtion

To address the extreme class imbalance in our datasets (positive-to-negative ratios of 1:19 and 1:15), we employed an asymmetric loss function (ASL) for model optimization. As demonstrated in the Figure 2C, ASL significantly outperformed both focal loss and binary cross-entropy (BCE) loss across all key metrics. In the *E. coli* dataset, while maintaining high accuracy and specificity, along with superior AUC (0.918) and AUPR (0.794) scores. Similarly, for the *A. baumannii* dataset, ASL demonstrated outstanding performance with substantially higher sensitivity and F1-score compared to other loss functions, while attaining

AUC and AUPR values of 0.781 and 0.400, respectively. Observed that models trained with BCE and Focal Loss exhibit poor specificity, sensitivity, and precision on negative samples when used with the capsule network. Under severe class imbalance, ASL maintains high overall accuracy while achieving substantially better sensitivity and precision; its AUC and AUPR are also acceptable.These findings strongly validate that the asymmetric loss function can effectively enhance the model's ability to recognize minority classes in extremely imbalanced datasets, preventing the model from collapsing into trivial majority-class-only predictions, thereby delivering more comprehensive and balanced performance.

### 4.4 Performance of different feature extraction methods of molecules

Our study systematically evaluated nine distinct molecular feature extraction approaches, including: (1) five individual molecular encoders (ChemBERTa-2, Smole-BERT, Molformer, MoleBERT, and MolCLR), (2) their simple concatenated combination, and (3) cross-attention optimized feature representations.

The cross-attention fusion module proposed in this study demonstrates significant advantages in Figure 2D. Detailed evaluation metrics are in Table S4. Comparative analysis of different molecular encoding methods reveals that the cross-attention enhanced features outperform other encoding approaches across both datasets. In the *E. coli* dataset, MoleBERT achieved the highest sensitivity but exhibited relatively low precision, indicating a higher false-positive rate when this feature set is used in isolation. In both datasets, ChemBERTa and MolCLR showed high specificity but low sensitivity and precision, suggesting that relying on either feature alone is inadvisable. Compared with simple feature concatenation, our model attains superior performance across all evaluated metrics, indicating that the cross-attention fusion module makes a positive contribution. Specifically, while simple feature concatenation achieves perfect specificity and precision in the *E.coli* dataset, its lower sensitivity and F1-score indicate inadequate recognition capability for rare positive samples. The cross-attention module effectively addresses this limitation, improving sensitivity to 0.667 and F1-score to 0.762 while maintaining high specificity. For the *A.baumannii* dataset, although the cross-attention processed features show slightly lower sensitivity compared to concatenated features, they achieve higher AUPR and F1-score with comparable AUC scores. These results conclusively demonstrate that the cross-attention module can more effectively integrate multi-source molecular features, significantly enhancing the recognition of scarce positive samples while maintaining overall model performance.

### 4.5 Feature analysis

Discriminative features are crucial for building robust computational classifiers. In this study, we employed GradientExplainer from SHAP analysis[44, 45] to examine 100 sample data points, revealing distinct contribution patterns of molecular features to model predictions. As illustrated in Figures 4A and 4B for the *E. coli* dataset, MolFormer features demonstrated the most significant impact on prediction outcomes, followed by Smole-BERT features.

Importantly, comparative analysis between Figures 3A and 3C highlights that the specific contribution of identical features varies substantially across different datasets.

## 4.6 Case Study

In this study, we used a classification dataset constructed by rigorously screening compounds from the AntiBac-Pred database in the study by Pogodin et al.[46]. For predicting antibacterial activity against *E.coli* and *A.baumannii.* A binary classification system was established based on minimum inhibitory concentration (MIC) values, using 10000nM as the threshold. Notably, our analysis specifically included compounds from the CHEMBL354 (for *E.coli*) and CHEMBL614425 (for *A.baumannii*) series, while explicitly excluding their resistant counterparts (CHEMBL354-R and CHEMBL614425-R).

We evaluated model performance using a validation set of 10 randomly selected compounds. Detailed information is shown in Table S7. For the *E.coli* dataset, the model demonstrated excellent predictive capability, achieving an accuracy of 0.8 and AUPR of 0.920, indicating its effectiveness in identifying antimicrobial compounds with *E. coli* activity. However, predictive performance was comparatively weaker for the *A.baumannii* dataset, with an accuracy of 0.5 and more frequent false positives, although the AUPR reached 0.641, suggesting room for improvement in predicting *A. baumannii* related antimicrobial compounds. This performance disparity likely stems from inherent differences in compound structural features or activity distributions between the two datasets, highlighting the need for future research to develop more specific feature extraction methods or model optimization strategies tailored for *A. baumannii*. The results underscore the importance of pathogen-specific model adaptation in antimicrobial activity prediction.

In addition to the compounds with explicitly reported antibacterial activity against *E.coli* and *A.baumannii*, we also selected newly discovered antibiotics and therapeutic agents for bacterial infections from recent studies to evaluate the generalizability of our model.

The study by Muteeb.G et al.[47] reported five novel antibiotics targeting drug-resistant bacteria: Teixobactin, Lefamulin, Zoliflodacin, Cefiderocol, and Eravacycline. Tangden.T[48] et al. identified four additional β-lactam antibiotics that have successfully entered the market and are active against carbapenem-resistant Gram-negative bacteria (CR-GNB) pathogens listed as “critical priority” by the World Health Organization (WHO), including cerftolozane/tazobactam, ceftazidime/avibactam, meropenem/varborbactam and imipenem.

The mechanisms of action of these antibiotics have been well-documented, with most exhibiting antibacterial activity against Gram-negative bacteria. Based on this feature, we obtained the SMILES notations of these compounds and input them into our model for prediction. Detailed information is shown in Table 4.

**Table 4.** Compounds selected based on recent research literature on newly developed antibiotics with antibacterial activity against *E.coli* or *A.baumannii* along with their model prediction results.

| **PubChem ID** | **Name** | **Antibacterial against *E.coli*[a]** | **Non-Antimicrobial Possibility on *E.coli* model** | **Antimicrobial Possibility on *E.coli* model** | **Antibacterial against *A.baumannii*[a]** | **Non-Antimicrobial Possibility on *A.baumannii* model** | **Antimicrobial Possibility on *A.baumannii* model** | **Mechanism of Action** | **Ref** |
|---|---|---|---|---|---|---|---|---|---|
| 86341926 | Teixobactin | 0 | 0.14199674 | 0.67199826 | 0 | 0.94832280 | 0.34149100 | Binds to lipid II's pyrophosphate-sugar moiety, forming supramolecular fibrils that disrupt membrane integrity | [49] |
| 76685216 | Zoliflodacin | 1 | 0.11834149 | 0.63064330 | 1 | 0.95306516 | 0.29162323 | Inhibits DNA gyrase/topoisomerase IV, blocking DNA replication and preventing religation of cleaved DNA | [50] |
| 77843966 | Cefiderocol | 1 | 0.17545433 | 0.67352563 | 1 | 0.93978600 | 0.49586460 | Acts as a siderophore-cephalosporin, hijacking bacterial iron transport to enter cells and bind PBP3 for cell wall inhibition | [51] |
| 56951485 | Eravacycline | 1 | 0.13054182 | 0.66077120 | 1 | 0.96027170 | 0.24948232 | Binds the 30S ribosomal subunit to halt protein | [52] |
| | | | | | | | | synthesis | |
| 58076382 | Lefamulin | 0 | 0.16541544 | 0.66996443 | 0 | 0.95985144 | 0.28726515 | Targets the 50S subunit's peptidyl transferase center to block translation | [53] |
| 104838 | Imipenem | 1 | 0.12806416 | 0.67288715 | 0 | 0.94260470 | 0.40595960 | Inhibits cell wall synthesis while resisting β-lactamase degradation. | [57] |

[a]: 1means active while 0 means inactive.

Through a systematic review of relevant literature, we have accurately obtained information on the antibacterial activity of a series of compounds against either *E.coli* or *A.baumannii*. For each compound, the literature elaborates in detail on the mechanisms by which it exerts its antibacterial effects[46, 48]. For instance, Teixobactin can interfere with the synthesis of bacterial cell walls, while Cefiderocol can bind to key enzymes involved in bacterial cell wall synthesis, hindering the cross-linking of cell wall peptidoglycans[49, 50]. This leads to damage to the cell wall structure, and the bacteria eventually rupture and die due to their inability to withstand intracellular osmotic pressure.

For the successfully predicted compounds with antibacterial activity, we have conducted in-depth research to identify their corresponding biological enzymes. Taking Zoliflodacin as an example, its mechanism of action involves inhibiting DNA gyrase/topoisomerase IV, blocking DNA replication, and preventing the religation of cleaved DNA strands. The literature mentions that Zoliflodacin can induce the SOS response to DNA damage in *E.coli* at levels similar to those induced by ciprofloxacin[51].

To further verify the prediction results, we employed molecular docking technology to perform docking simulations between these compounds and their corresponding biological enzymes, and observed the magnitude of binding energies. Binding energy is an important indicator for measuring the strength of interactions between compounds and biological enzymes; a lower binding energy indicates that the two can form a more stable complex, thereby exerting a more effective antibacterial effect. For the enzymes docked with the compounds, as well as the binding energies obtained from the docking of these enzymes with the compounds, details are shown in Figure 3E.

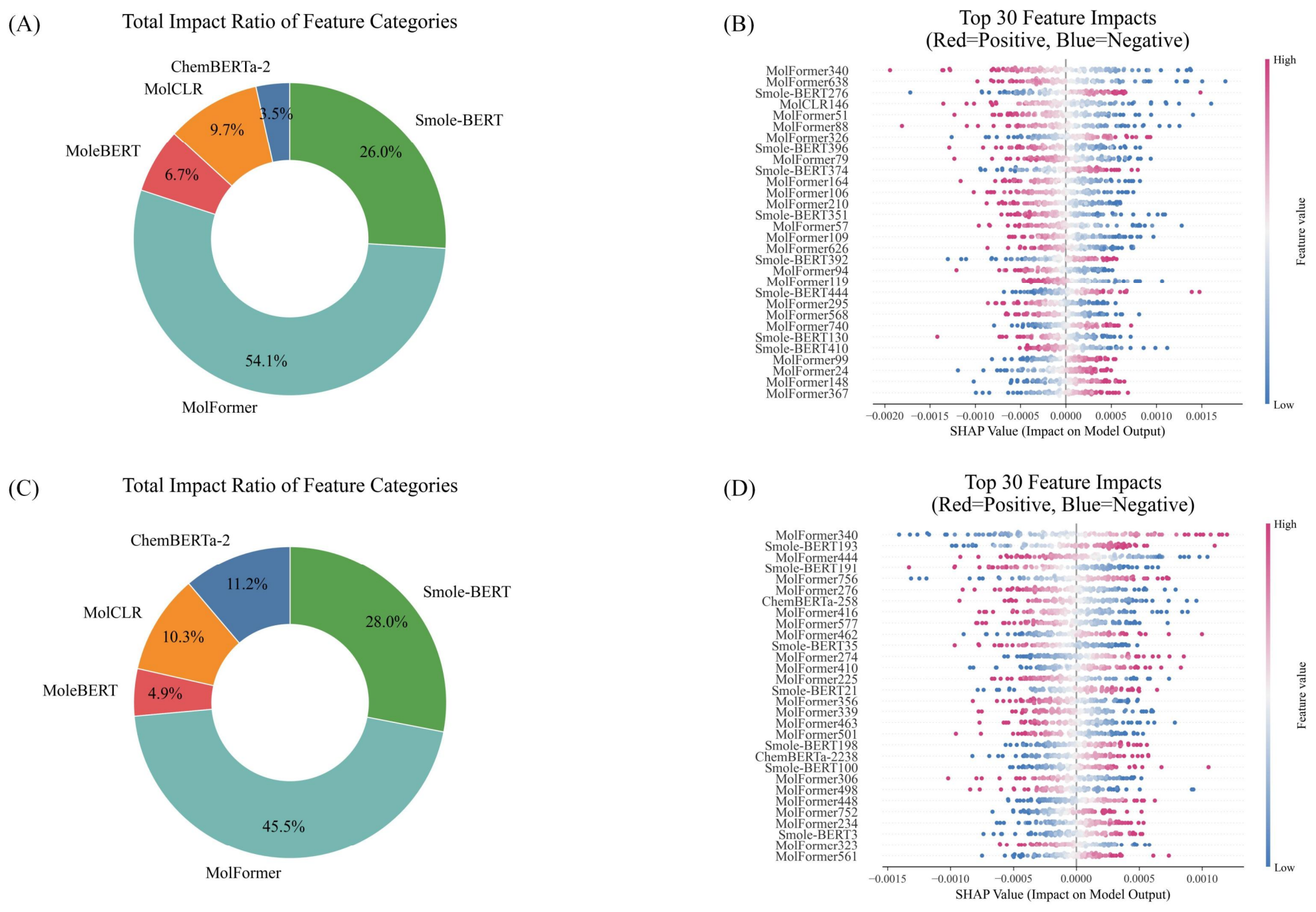

(E)

| Compounds | Target (PDB Entry) | Calculated affinity (kcal/mol) |
|---|---|---|
| Zoliflodacin | DNA gyrase (6RKS) | -6.813 |
| Cefiderocol | β-lactamases (1ZG4) | -6.825 |
| Eravacycline | 30S ribosomal subunit (7OE1) | -5.443 |
| Imipenem | PBPa1 (2C6W) | -5.701 |
| | PBP2 (6G9P) | -7.29 |

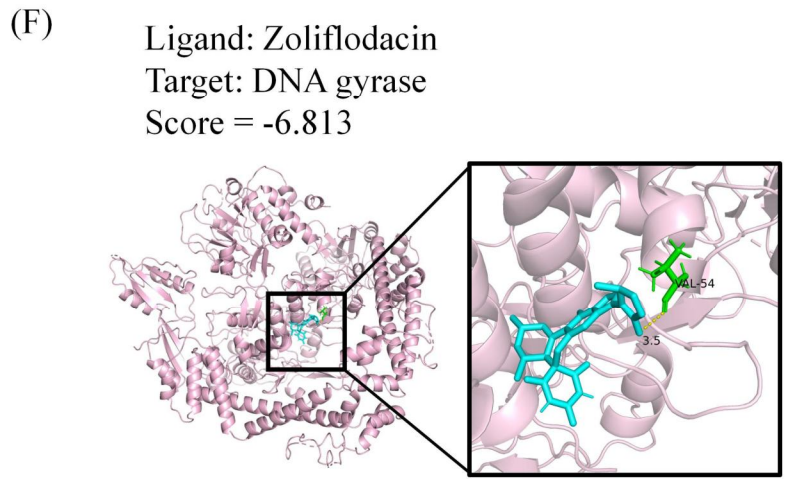

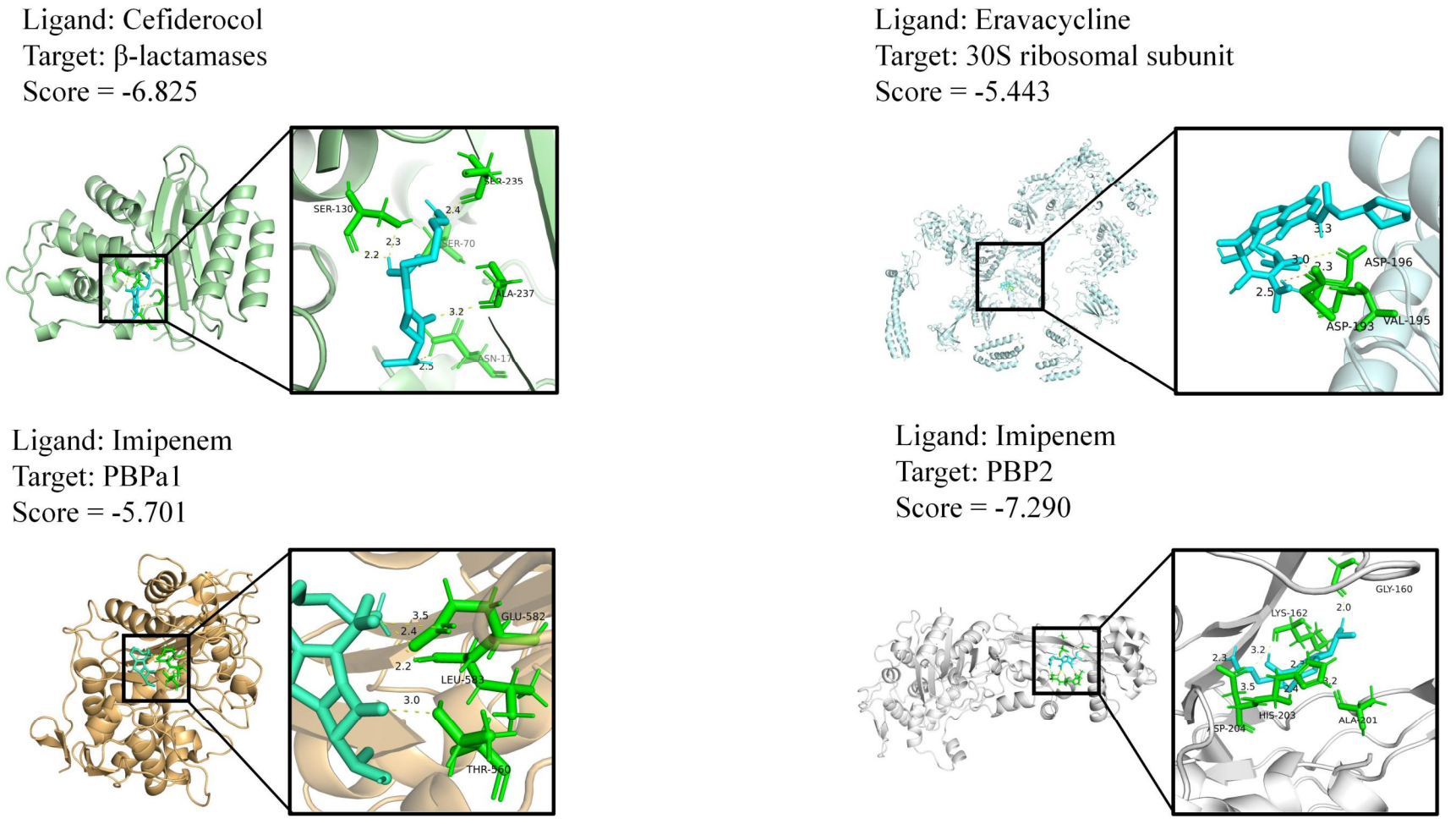

**Figure 3.** Integrated feature analysis and molecular docking results. SHAP-based feature importance and dependence visualizations for concatenated data, and case-study docking results for selected compounds.

Panels: (A-B) SHAP feature importance and top 30 most important features for *E. coli*; (C-D) SHAP feature importance and top 30 most important features for *A. baumannii*; (E) Summary of molecular docking targets and docking scores for compounds with confirmed antibacterial activity and correctly predicted by the model; (F) Representative docking poses of selected drugs into candidate target proteins in *E. coli*, with hydrogen bond interactions annotated.

The species listed in the table are all strains of *E. coli*. This selection was made because, in the model trained specifically on the *A.baumannii* dataset, we are able to effectively identify drugs lacking antibacterial effects; however, our primary emphasis lies in precisely predicting those with antibacterial activity. Teixobactin acts on lipids instead of proteins. Lefamulin predominantly targets pathogens responsible for community-acquired pneumonia, such as *Streptococcus pneumoniae*, and demonstrates limited efficacy against Gram-negative bacteria. Hence, these two compounds were excluded from consideration. Drawing from existing research, we identified the relevant targets for the compounds in the Protein Data Bank (PDB)[52]. Using PyMOL, we filtered out non-protein small molecules and subsequently conducted molecular docking with Autodock Vina[53, 54]. Detailed docking parameters can be found in Table S7. Upon completion of the docking simulations, we obtained the calculated binding affinities and, using PyMOL, identified the residues that exhibited the lowest ΔG values and formed interactions with the target protein. This case study effectively demonstrates the robustness and enhanced performance of our proposed model under real-world conditions.

## 4.7 Web Server

We developed a web interface (https://awi.cuhk.edu.cn/~AIDD/CapMolPred/indexpage.php) to facilitate rapid prediction of chemical compound antimicrobial activity (Fig. 4). The interface accepts multiple input formats—canonical SMILES, drug name, or PubChem Compound ID—and performs automated preprocessing and model inference, with a target turnaround time of approximately one minute to ensure a responsive, user-friendly experience. Prediction outputs comprise binary inhibitory labels and associated probabilities for *Escherichia coli* and *Acinetobacter baumannii*, providing an interpretable assessment of each compound's predicted antimicrobial potential. To promote transparency and enable downstream analyses, the interface includes a download section exposing the datasets employed during prediction; these are provided in CSV format and contain compound identifiers, canonical SMILES, assay metadata, and activity annotations for *E. coli* and *A. baumannii.*

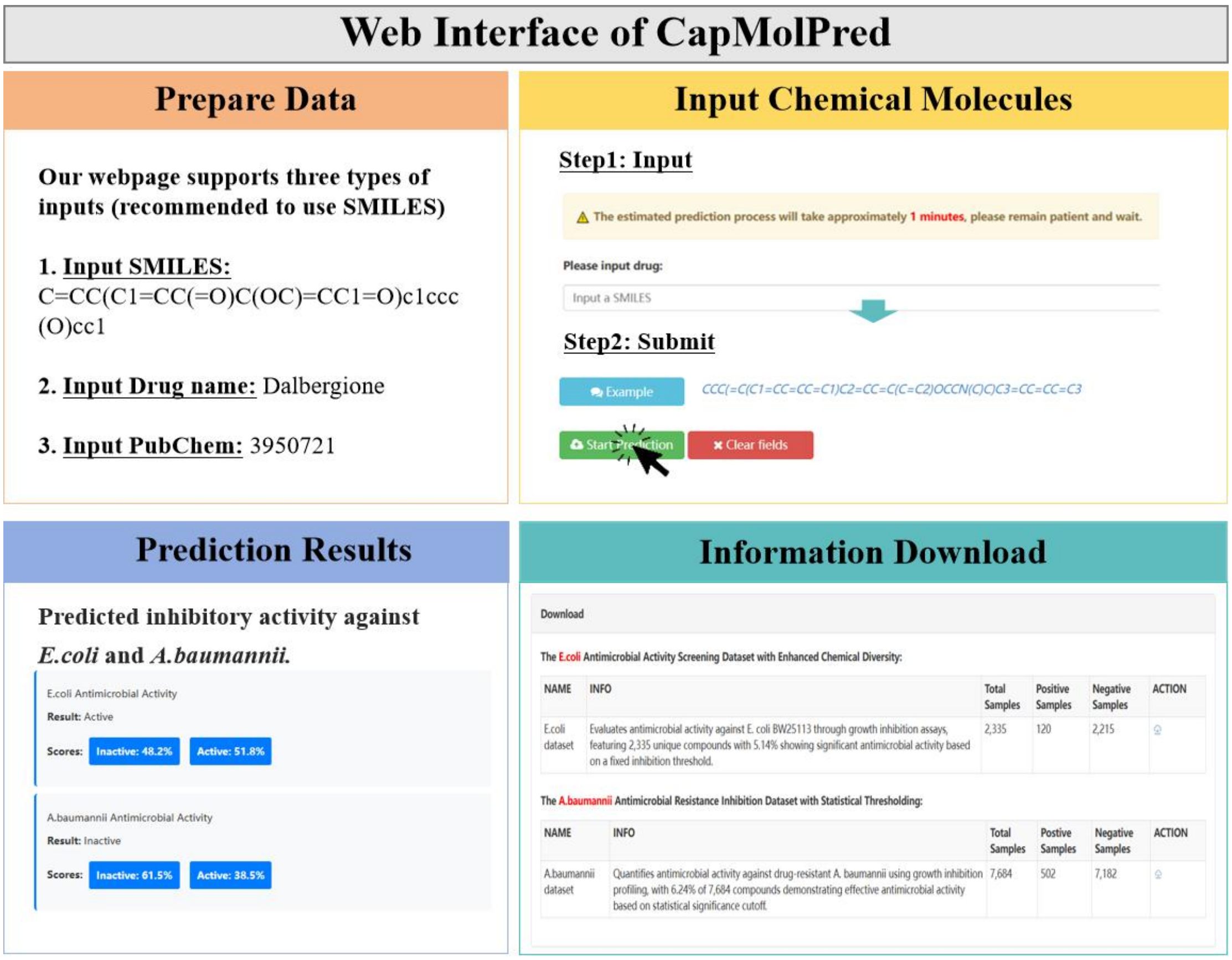

**Figure 4.** Workflow of our web interface, comprising four stages: Data preparation, Chemical Molecule Input, Prediction Results, and Information Download.

## 5 Conclusion and Discussion

Molecules exhibiting antimicrobial activity play vital roles in combating pathogenic microorganisms and are integral to a wide range of biological and clinical applications. In therapeutic development, specific small molecules can serve as potent antimicrobial agents, either inhibiting pathogen growth or eradicating infectious organisms altogether. Therefore, accurate prediction of the antimicrobial activity of molecular compounds is of paramount importance for drug discovery and the advancement of novel antimicrobial therapies.

In this study, we propose a deep learning-based model aimed at predicting the antibacterial activity of small-molecule drugs. We extract features from small molecules' SMILES representations using ChemBERTa-2, Smole-BERT, MolFormer, MoleBERT, and MolCLR, and concatenate the molecular embeddings obtained from these models. The concatenated embeddings are then refined through a cross-attention module to optimize feature fusion. Furthermore, our model incorporates a capsule network architecture as the classifier, which categorizes positive and negative samples based on hierarchical feature representations. Compared to various baselines and previous models, our model has demonstrated robust performance as a sequence-based approach for predicting potential antibacterial compounds.

The reliability of our model arises from several key factors. First, we utilize BERT-based pretrained models to extract features from small molecules' SMILES sequences; these models exhibit excellent chemical sequence encoding capabilities due to extensive pretraining on large datasets. Second, we effectively handle multimodal embeddings and address the class imbalance inherent in the datasets. Third, we employ a next-generation neural network, the capsule network, as the classifier to enhance classification performance. Finally, we conducted predictions on the newly discovered antibiotics, obtained accurate results, and then carried out case studies on molecular docking to further demonstrate the superiority and reliability of our model.

Despite significant progress, current research still faces several critical challenges. At the data level, issues such as class imbalance, scarcity of labeled data for emerging pathogens, and the consequent risk of overfitting necessitate more effective regularization methods and data augmentation strategies. As demonstrated by Mongia et al., modeling antibacterial activity as a continuous variable—such as inhibition rate—through regression rather than binary classification can better leverage existing data, although this approach demands larger training datasets. Beyond the loss function optimization employed in this study and the commonly used oversampling and undersampling techniques, constructing more comprehensive and balanced datasets remains essential. Moreover, current models generally lack explicit recognition of specific pharmacophores (e.g., β-lactam ring) or inhibitory substructures, which may be critical determinants of antibacterial mechanisms. The method proposed herein can be further improved by incorporating more detailed chemical space representations and factors vital to antibiotic activity, such as serum protein binding[55]. Through iterative data generation, model retraining, and substructure identification, more complete chemical space characterizations can be developed, enhancing the identification and classification of promising candidates. Additionally, some studies have validated machine learning-predicted antibacterial compounds via wet-lab experiments, thereby reinforcing model reliability. For instance, Swanson et al. developed a novel generative AI model for small molecule drug design, discovering six structurally novel molecules (with a hit rate of approximately 10%) active against *A.baumannii* and other phylogenetically distinct ESKAPE pathogens [56]. This underscores a key direction for future work: integrating computational prediction with experimental validation to systematically enhance model reliability and translational value.

## FIGURE AND TABLE LEGENDS

**Table 1.** The details of existing computational methods.

**Table 2.** Statistics of sample distribution in *E.coli* and *A.baumannii* datasets.

**Table 3.** Compounds selected based on recent research literature on newly developed antibiotics with antibacterial activity against *E.coli* or *A.baumannii* along with their model prediction results.

**Figure 1.** This figure illustrates the workflow and application of our research.**(A)** Dataset statistics and binary label generation for *Escherichia coli* and *Acinetobacter baumannii*.

**(B)** Workflow illustrating multi-modal feature extraction, subsequent cross attention module, and final prediction utilize a capsule-network predictor with dynamic routing.

**(C)** Performance evaluation using standard classification metrics.

**(D)** Comparative analysis against baseline and state-of-the-art models.

**(E)** Feature importance interpretation via SHAP analysis.

**(F)** Case study using newly reported antibiotics compiled from published literature.

**(G)** The model is deployed on a web server, providing a user-friendly interface for predicting antimicrobial activity of molecules.

**Figure 2.** Consolidated comparison of model performance and feature-space visualization across *E. coli* and *A. baumannii* datasets. Panels: (A) Comparative binary-classification performance of our proposed model versus existing methods on *E. coli* and *A. baumannii*; (B) Ablation study of different network classifier for both datasets; (C) Comparative performance using different loss functions for both datasets; (D) Performance across alternative feature representations for both datasets; (E-F) t-SNE visualizations of selected CapsNet layers for *E. coli* (E) and *A. baumannii* (F), with points colored by class label.

**Figure 3.** Integrated feature analysis and molecular docking results. SHAP-based feature importance and dependence visualizations for concatenated data, and case-study docking results for selected compounds.

Panels: (A-B) SHAP feature importance and top 30 most important features for *E. coli*; (C-D) SHAP feature importance and top 30 most important features for *A. baumannii*; (E) Summary of molecular docking targets and docking scores for compounds with confirmed antibacterial activity and correctly predicted by the model; (F) Representative docking poses of selected drugs into candidate target proteins in *E. coli*, with hydrogen bonds interactions annotated.

**Figure 4.** The workflow of the our web interface.

**Table 1.** The details of existing computational methods.

| Tool | Journal | Dataset | Method Class | Citation | Model Construction | Limitation | Web Interface | Code Available |
|---|---|---|---|---|---|---|---|---|
| Chemprop | Cell (2020) | *E.coli* | Deep Learning (Directed Message Passing Neural Network) | 2454 | A Directed Message Passing Neural Network (D-MPNN) was trained on 2,335 molecules with growth inhibition data, using molecular graph representations as input and applying transfer learning from a broad-spectrum prediction task. | 1. Requires experimental validation for clinical relevance<br>2. Model interpretability challenges for complex molecules | × | https://github.com/chemprop/chemprop |
| Chemprop | *Nature Chemical Biology (2023)* | *A.baumannii* | Deep Learning (Directed Message Passing Neural Network) | 268 | A D-MPNN architecture was fine-tuned for A. baumannii using dual-path inputs—molecular graphs and precomputed features—and trained it on 7,684 experimentally tested compounds, leveraging transfer learning from a broad-spectrum antibiotic model. | 1. Limited training data for narrow-spectrum antibiotics<br>2. Computational screening bias toward synthetic feasibility | × | https://github.com/chemprop/chemprop |
| InterPred | Science Advances (2019) | *E.coli* & *A.baumannii* | Interpretable Machine Learning (Multi-task Classification) | 13 | Extra trees/logistic regression classifier with L1 regularization is trained on bacterial morphological features from Cell Painting data to predict five classes of mechanism of action (DNA, RNA, protein, cell wall, and membrane targets), leveraging transfer learning from antibiotic resistance data. | 1. Limited to Gram-negative bacteria *(E. coli* focus)<br>2. Morphological features require high-content imaging<br>3. Cannot detect novel MoA outside training classes | × | https://gitlab.com/mongolicious/interpretableml-for-mechanism-of-action/ |

| | | | | | | | |
|---|---|---|---|---|---|---|---|
| MFAGCN | Scientific Reports (2024) | Machine Learning (Ensemble Learning ) | 1 | Molecular descriptors (209-D) were calculated using RDKit, followed by feature selection via Recursive Feature Elimination (RFE). An XGBoost classifier was trained on the PubChem antimicrobial activity dataset containing over 12,000 compounds. | 1. Dependency on existing activity labels (potential labeling bias)<br>2. Limited generalizability to novel structural domains | × | https://github.com/MFAGCN/MFAGCN. |
| our model | - | Capsule Network | - | A multi-modal model was constructed by extracting features from molecular representations using both 1D sequence encoders (ChemBERTa-2, SmoleBERT, MolFormer) and 2D graph encoders (MoleBERT, MolCLR). These features were fused through a cross-attention module combined with concatenation, and the final prediction was made using a capsule network architecture with dynamic routing. | - | √ | - |

*The citation counts in this table were retrieved on Jul 15, 2025 from Google Scholar.

**Table 2.** Statistics of sample distribution in *E.coli* and *A.baumannii* datasets.

| **Dataset** | **Positive sample** | **Negative sample** | **Total** |
|---|---|---|---|
| *E.coli* | 120 | 2215 | 2335 |
| *A.baumannii* | 502 | 7182 | 7684 |

**Table 3.** Compounds selected based on recent research literature on newly developed antibiotics with antibacterial activity against *E.coli* or *A.baumannii* along with their model prediction results.

| PubChem ID | Name | Antibacterial against *E.coli*[a] | Non-Antimicrobial Possibility on *E.coli* model | Antimicrobial Possibility on *E.coli* model | Antibacterial against *A.baumannii*[a] | Non-Antimicrobial Possibility on *A.baumannii* model | Antimicrobial Possibility on *A.baumannii* model | Mechanism of Action | Ref |
|---|---|---|---|---|---|---|---|---|---|
| 86341926 | Teixobactin | 0 | 0.14199674 | 0.67199826 | 0 | 0.94832280 | 0.34149100 | Binds to lipid II's pyrophosphate-sugar moiety, forming supramolecular fibrils that disrupt membrane integrity | [49] |
| 76685216 | Zoliflodacin | 1 | 0.11834149 | 0.63064330 | 0 | 0.95306516 | 0.29162323 | Inhibits DNA gyrase/topoisomerase IV, blocking DNA replication and preventing religation of cleaved DNA | [50] |
| 77843966 | Cefiderocol | 1 | 0.17545433 | 0.67352563 | 1 | 0.93978600 | 0.49586460 | Acts as a siderophore-cephalosporin, hijacking bacterial iron transport to enter cells and bind PBP3 for cell wall inhibition | [51] |
| 56951485 | Eravacycline | 1 | 0.13054182 | 0.66077120 | 1 | 0.96027170 | 0.24948232 | Binds the 30S ribosomal subunit to halt protein | [52] |
| | | | | | | | | synthesis | |
| 58076382 | Lefamulin | 0 | 0.16541544 | 0.66996443 | 0 | 0.95985144 | 0.28726515 | Targets the 50S subunit's peptidyl transferase center to block translation | [53] |
| 104838 | Imipenem | 1 | 0.12806416 | 0.67288715 | 0 | 0.94260470 | 0.40595960 | Inhibits cell wall synthesis while resisting β-lactamase degradation. | [57] |

[a]: 1means active while 0 means inactive.

**Figure 1.** This figure illustrates the workflow and application of our research.**(A)** Dataset statistics and binary label generation for *Escherichia coli* and *Acinetobacter baumannii*.
**(B)** Molecular representation using SMILES strings and RDKit-derived graph descriptors.
**(C)** Multi-modal feature extraction:

· **1D Sequence Encoders:** ChemBERTa-2, SmoleBERT, MolFormer

· **2D Graph Encoders:** MoleBERT, MolCLR

· **Feature Fusion:** Cross-attention module with concatenation

· **Prediction Model:** Capsule network architecture with dynamic routing

**(D)** Performance evaluation using standard classification metrics.
**(E)** Comparative analysis against baseline and state-of-the-art models.
**(F)** Feature importance interpretation via SHAP analysis.
**(G)** Web server deployment for user-friendly antibacterial property prediction.

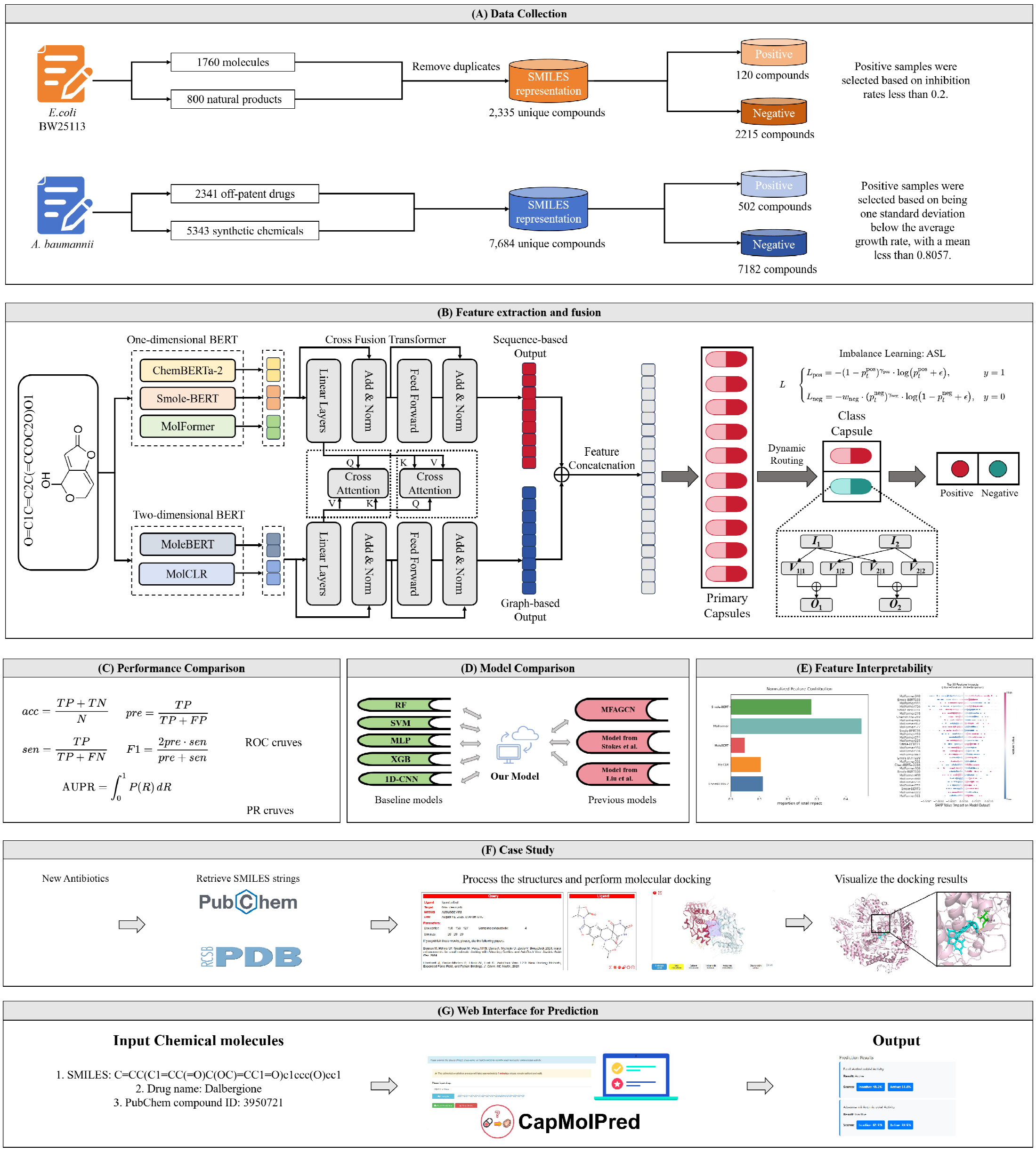

**Figure 2.** Consolidated comparison of model performance and feature-space visualization across *E. coli* and *A. baumannii* datasets. Panels: (A) Comparative binary-classification performance of our proposed model versus existing methods on *E. coli* and *A. baumannii*; (B) Ablation study of different network classifier for both datasets; (C) Comparative performance using different loss functions for both datasets; (D) Performance across alternative feature representations for both datasets; (E-F) t-SNE visualizations of selected CapsNet layers for *E. coli* (E) and *A. baumannii* (F), with points colored by class label.

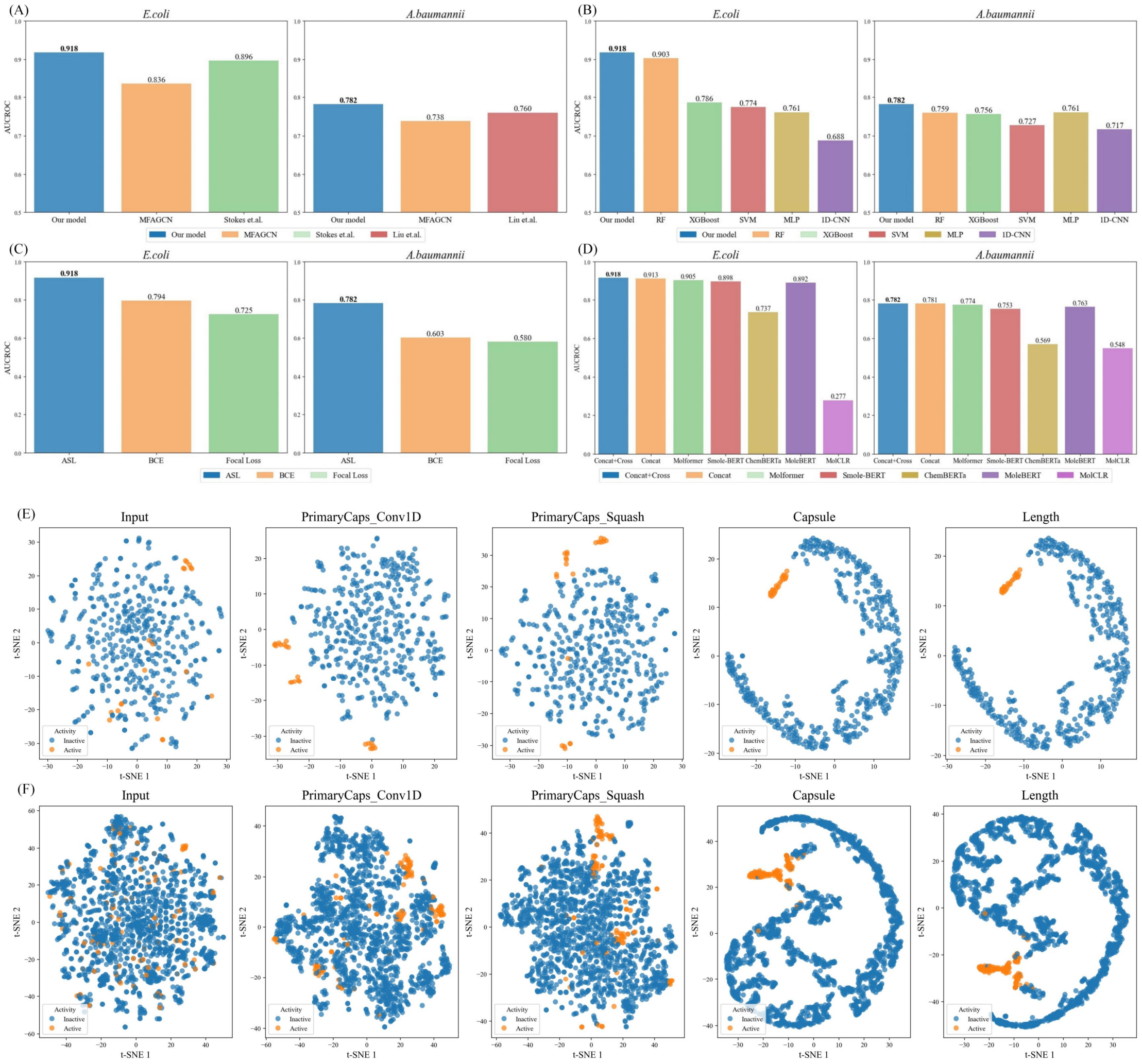

**Figure 3.** Integrated feature analysis and molecular docking results. SHAP-based feature importance and dependence visualizations for concatenated data, and case-study docking results for selected compounds.

Panels: (A-B) SHAP feature importance and top 30 most important features for *E. coli*; (C-D) SHAP feature importance and top 30 most important features for *A. baumannii*; (E) Summary of molecular docking targets and docking scores for compounds with confirmed antibacterial activity and correctly predicted by the model; (F) Representative docking poses of selected drugs into candidate target proteins in *E. coli*, with hydrogen bonds interactions annotated.

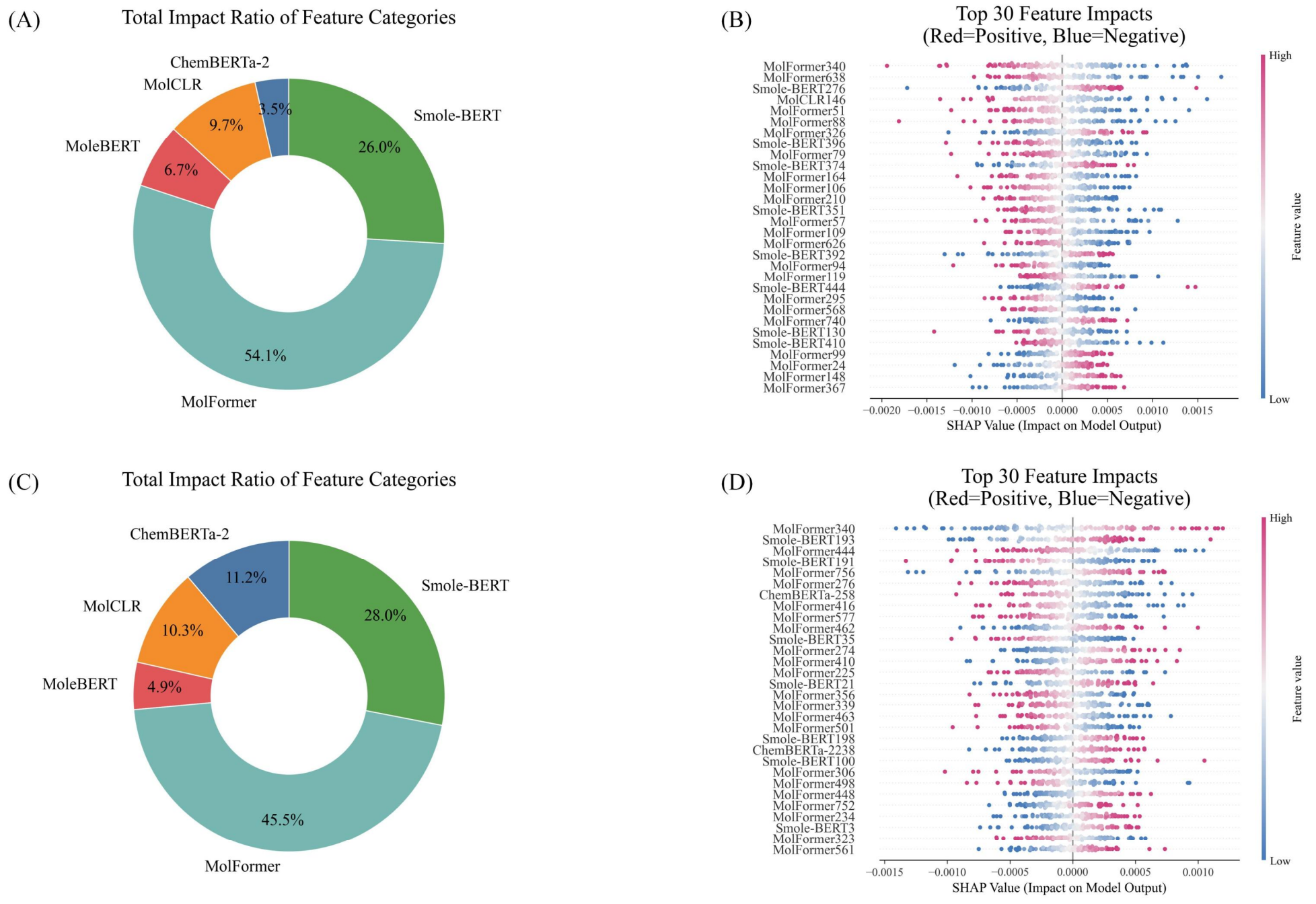

(E)

| Compounds | Target (PDB Entry) | Calculated affinity (kcal/mol) |
|---|---|---|
| Zoliflodacin | DNA gyrase (6RKS) | -6.813 |
| Cefiderocol | β-lactamases (1ZG4) | -6.825 |
| Eravacycline | 30S ribosomal subunit (7OE1) | -5.443 |
| Imipenem | PBPa1 (2C6W) | -5.701 |
| | PBP2 (6G9P) | -7.29 |

(F)

Ligand: Zoliflodacin
Target: DNA gyrase
Score = -6.813

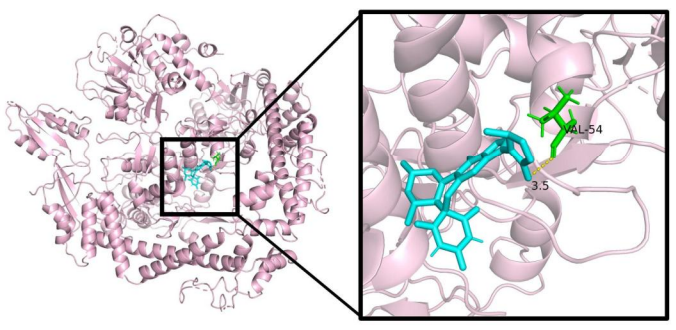

Ligand: Cefiderocol
Target: β-lactamases
Score = -6.825

Ligand: Eravacycline
Target: 30S ribosomal subunit
Score = -5.443

Ligand: Imipenem
Target: PBPa1
Score = -5.701

Ligand: Imipenem
Target: PBP2
Score = -7.290

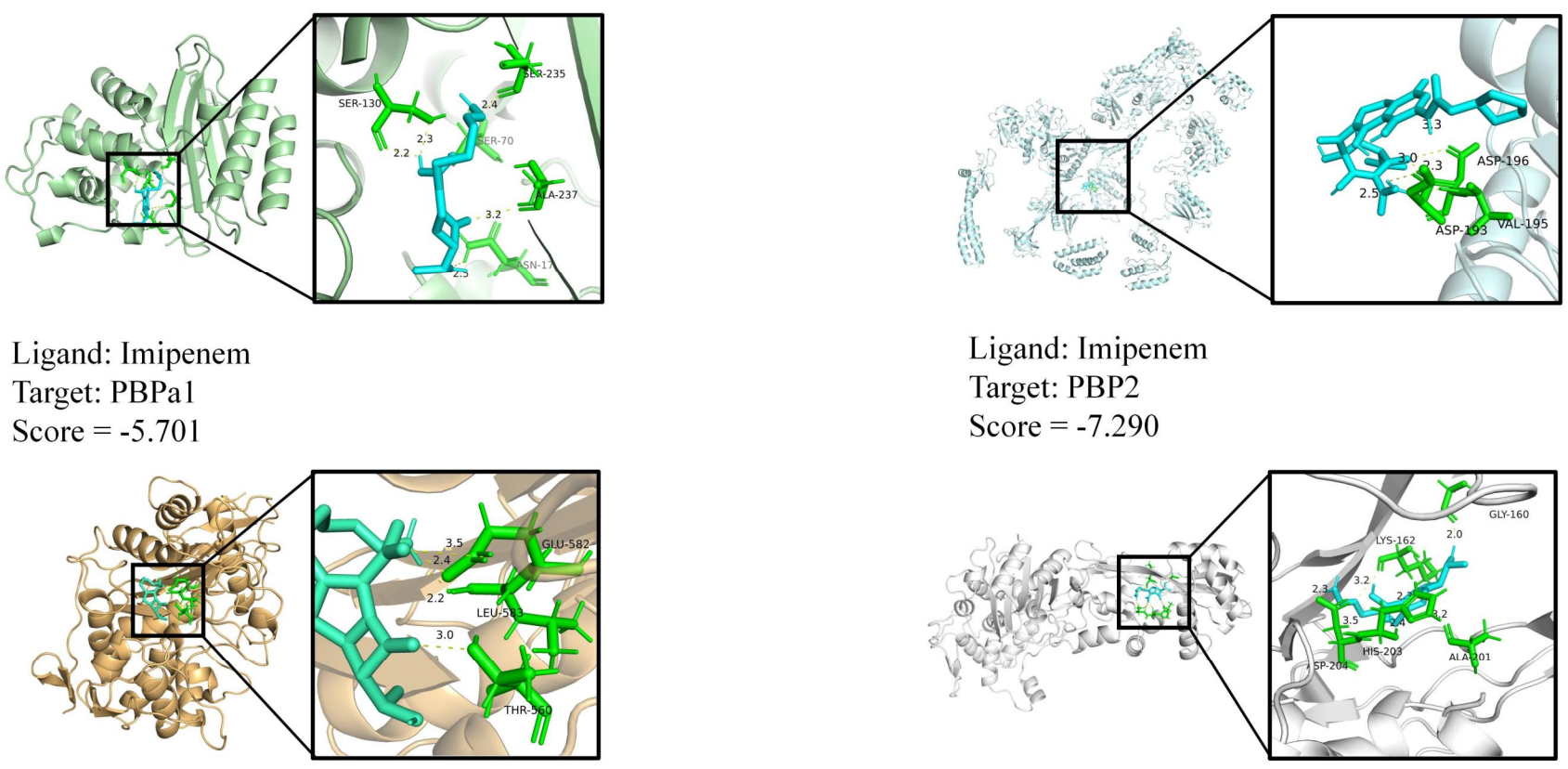

**Figure 4.** Workflow of our web interface, comprising four stages: Data preparation, Chemical Molecule Input, Prediction Results, and Information Download.

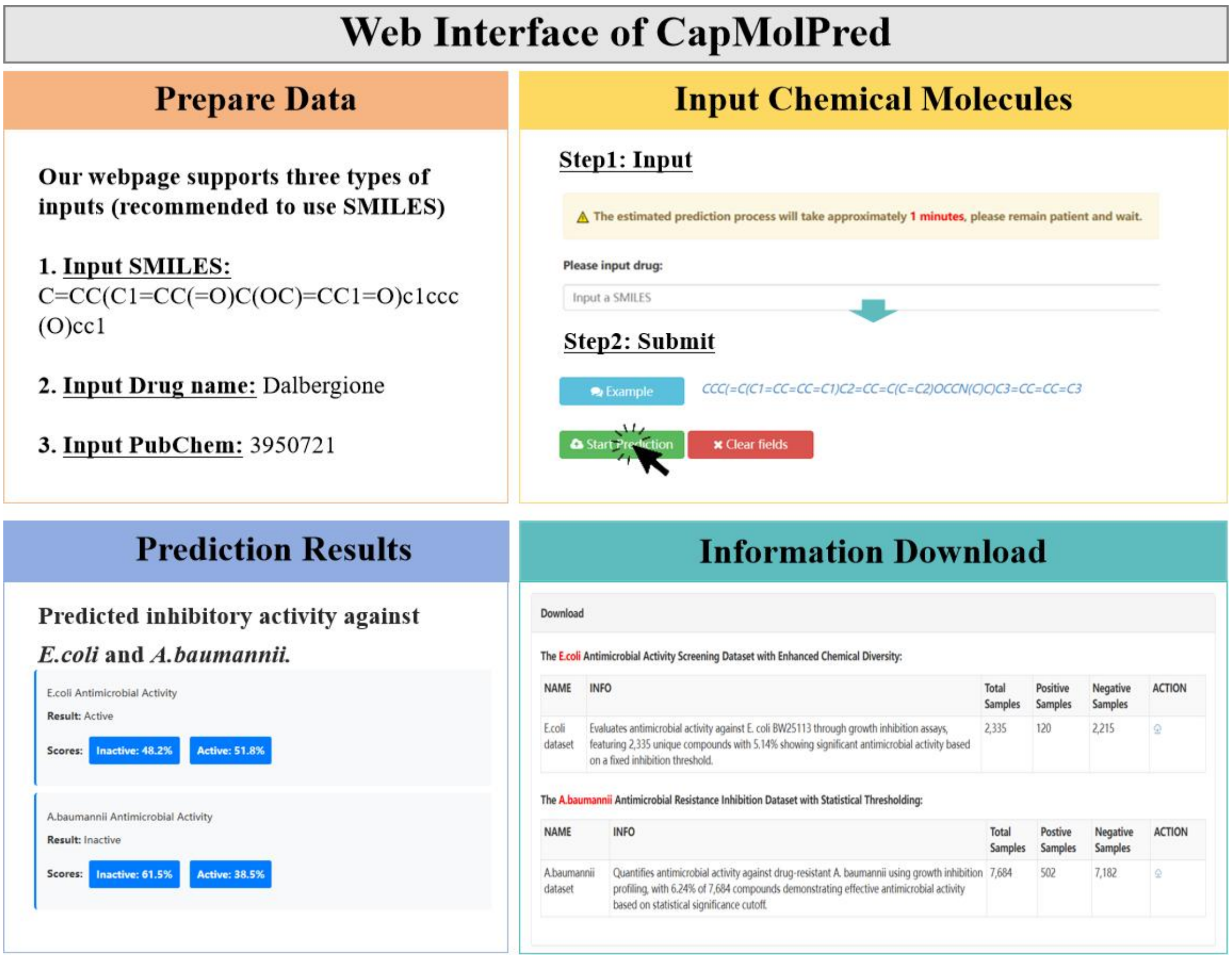